\documentclass[aps,prl,amsmath,amssymb,twocolumn,floatfix]{revtex4-1}
\usepackage{graphicx}
\usepackage{dcolumn}
\usepackage{bm}
\usepackage{rotating}
\usepackage{epsfig}
\usepackage{graphicx}
\usepackage{epstopdf}
\usepackage{amsmath}

\begin{document}

\title{Coexistence of ferromagnetic fluctuations and superconductivity in the actinide superconductor UTe$_2$}

\author{Shyam~Sundar,$^1$, S.~Gheidi,$^1$ K.~Akintola,$^1$ A. M.~C\^{o}t\'{e},$^{1,2}$ S. R.~Dunsiger,$^{1,3}$ S. Ran,$^{4,5}$ N. P. Butch,$^{4,5}$ S. R.~Saha,$^5$ J.~Paglione,$^{5,6}$, and J. E.~Sonier$^{1,6}$}

\affiliation{$^1$Department of Physics, Simon Fraser University, Burnaby, British Columbia V5A 1S6, Canada \\
$^2$Kwantlen Polytechnic University, Richmond, British Columbia V6X 3X7, Canada \\
$^3$Centre for Molecular and Materials Science, TRIUMF, Vancouver, British Columbia V6T 2A3, Canada \\ 
$^4$NIST Center for Neutron Research, National Institute of Standards and Technology, Gaithersburg, MD 20899, USA \\
$^5$Center for Nanophysics and Advanced Materials, Department of Physics, University of Maryland, College Park, Maryland 20742, USA \\
$^6$Canadian Institute for Advanced Research, Toronto, Ontario M5G 1Z8, Canada}

\date{\today}
\begin{abstract}
We report low-temperature muon spin relaxation/rotation ($\mu$SR) measurements on single crystals of the actinide superconductor UTe$_2$.
Below 5~K we observe a continuous slowing down of magnetic fluctuations that persists through the superconducting (SC) transition
temperature ($T_c \! = \! 1.6$~K), but we find no evidence of long-range or local magnetic order down to 0.025~K. The temperature dependence of the dynamic 
relaxation rate down to 0.4~K agrees with the self-consistent renormalization theory of spin fluctuations for a three-dimensional weak itinerant ferromagnetic metal. 
Our $\mu$SR measurements also indicate that the superconductivity coexists with the magnetic fluctuations. 
\end{abstract}

\maketitle
The unusual physical properties of intermetallic uranium-based superconductors are primarily due to the U-$5f$ electrons having both localized and itinerant character. 
In a subclass of these compounds, superconductivity coexists with ferromagnetism. In URhGe and UCoGe \cite{Aoki:01,Huy:07} this occurs
at ambient pressure, whereas superconductivity appears over a limited pressure range in UGe$_2$ and UIr \cite{Saxena:00,Akazawa:04}.
With the exception of UIr, the Curie temperature of these ferromagnetic (FM) superconductors signficantly exceeds $T_c$, and the upper critical field 
$H_{c2}$ at low temperatures greatly exceeds the Pauli paramagnetic limiting field. These observations indicate that the SC phases in 
these materials are associated with spin-triplet Cooper pairing, and likely mediated by low-lying magnetic fluctuations in 
the FM phase \cite{Fay:80,Roussev:01,Kirkpatrick:01,Tateiwa:18}. The triplet state is specifically {\it non-unitary}, characterized by 
a non-zero spin-triplet Cooper pair magnetic moment due to alignment of the Cooper pair spins with the internal field generated by the pre-existing FM order.

Very recently, superconductivity has been observed in UTe$_2$ at ambient pressure below $T_c \! \sim \! 1.6$~K \cite{Ran:18}. 
The superconductivity in UTe$_2$ also seems to involve spin-triplet pairing, as evidenced by a strongly anisotropic critcial field $H_{c2}$ that exceeds 
the Pauli limit, and by the lack of any temperature dependence of the $^{125}$Te nuclear magnetic resonance (NMR) Knight shift through and below $T_c$. 
Furthermore, a large residual value of the Sommerfeld coefficient $\gamma$ is observed in the SC state, which is nearly 50~\% of the value 
of $\gamma$ above $T_c$ \cite{Ran:18,Aoki:19}. This suggests that only half of the electrons occupying states near the Fermi surface participate in 
spin-triplet pairing, while the remainder continue to form a Fermi liquid. While this is compatible with UTe$_2$ being a non-unitary spin-triplet superconductor 
(in which the spin of the Cooper pairs are aligned in a particular direction), unlike URhGe, UCoGe and UGe$_2$, there is no experimental evidence 
for ordering of the U-$5f$ electron spins prior to the onset of superconductivity. Instead, the normal-state $a$-axis magnetization exhibits scaling
behavior indicative of strong magnetic fluctuations associated with metallic FM quantum criticality \cite{Ran:18}.

Little is known about the nature of the magnetism in UTe$_2$ below $T_c$, including whether it competes or coexists with superconductivity. 
Specific heat measurements show no anomaly below $T_c$ \cite{Ran:18,Aoki:19}, but like other bulk properties may be insensitive to a FM transition with 
little associated entropy (such as small-moment itinerant ferromagnetism). NMR experiments indicate the development of low-frequency longitudinal magnetic 
fluctuations along the $a$-axis, but the corresponding NMR signal vanishes below 20~K \cite{Tokunaga:19}. Here we report $\mu$SR experiments 
on UTe$_2$ single crystals that confirm the absence of FM order below $T_c$ and demonstrate the presence of magnetic fluctuations consistent with 
FM quantum criticality that coexist with superconductivity. 

The UTe$_2$ single crystals were grown by a chemical vapor transport method. 
Powder x-ray diffraction (XRD), neutron scattering and Laue XRD
measurements indicate that the single crystals are of high quality. The details of the sample growth and characterization are given in Ref.~\cite{Ran:18}.
Zero-field (ZF), longitudinal-field (LF), transverse-field (TF), and weak transverse-field (wTF) $\mu$SR measurements were performed on a mosaic of 21 
single crystals. Measurements over the temperature range $0.02$~K~$\lesssim \! T \! \lesssim \! 5$~K were achieved using an Oxford Instruments 
top-loading dilution refrigerator on the M15 surface muon beam line at TRIUMF.
The UTe$_2$ single crystals covered $\sim \! 70$~\% of a 12.5~mm~$\times$~14~mm silver (Ag) sample holder. 
For the ZF-$\mu$SR experiments, stray external magnetic fields at the sample position were reduced to $\lesssim \! 20$~mG 
using the precession signal due to muonium (Mu $\equiv \! \mu^+$e$^-$) in intrinsic Si as a sensitive magnetometer \cite{Morris:03}.
The TF and LF measurements were performed with a magnetic field applied parallel to the linear momentum of the muon beam 
(which we define to be in the $z$-direction). The wTF experiments were done with the field 
applied perpendicular to the beam (defined to be the $x$-direction). The initial muon spin polarization {\bf P}(0) was directed parallel to the $z$-axis 
for the ZF, LF and wTF experiments, and rotated in the $x$-direction for the TF measurements. The $c$- or $a$-axis of the single crystals were
arbitarily aligned in the $z$-direction. All error bars herein denote uncertainties of one standard deviation.   

\begin{figure}
\centering
\includegraphics[width=8.5cm]{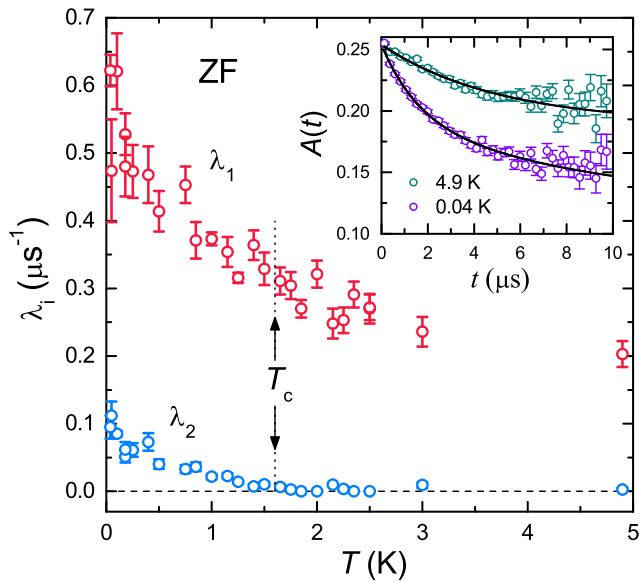}
\caption{(Color online) Temperature dependence of the ZF exponential relaxation rates obtained from fits of 
the ZF-$\mu$SR asymmetry spectra to Eq.~(\ref{eq:1}). Inset: Represenative ZF signals for $T \! = \! 4.9$~K and 
$T \! = \! 0.04$~K. The solid curves are the resultant fits to Eq.~(\ref{eq:1}).}
\label{fig1}
\end{figure}

Representative ZF-$\mu$SR asymmetry spectra for UTe$_2$ at $T \! = \! 0.04$~K and 4.9~K are shown in the inset of Fig.~\ref{fig1}. No oscillation indicative
of magnetic order is observed in any of the ZF-$\mu$SR spectra, which are well described by a three-component function consisting of two exponential 
relaxation terms plus a temperature-independent background term due to muons stopping outside the sample 
\begin{eqnarray} 
A(t) & = & A(0)P_z(t) \nonumber \\
& = &  A_1 e^{-\lambda_1 t} + A_2 e^{-\lambda_2 t} + A_{\rm B} e^{-\sigma^2 t^2} \, .
\label{eq:1}
\end{eqnarray}
The sum of the sample asymmetries $A_1 \! + \! A_2$ is a measure of the recorded decay events originating from muons stopping in the sample. 
A global fit of the ZF spectra for all temperatures assuming common values of the asymmetry parameters, yielded $A_1/A(0) \! = \! 24$~\%, 
$A_2/A(0) \! = \! 29$~\% and $A_{\rm B}/A(0) \! = \! 47$~\%. 
A previous $\mu$SR study of UGe$_2$ identified two muon stopping sites, with site populations of $\sim \! 45$~\% for one site and 
$\sim \! 55$~\% for the other \cite{Sakarya:10}, in excellent agreement with the results here. 
The temperature variation of the ZF relaxation rates $\lambda_1$ and $\lambda_2$ are shown in Fig.~\ref{fig1}.
The monotonic increase in $\lambda_1$ and $\lambda_2$ with decreasing temperature indicates that the local magnetic field sensed at each muon site is
dominated by a slowing down of magnetic fluctuations, as explained below. The difference in the size of the relaxation rates reflects a difference in the 
dipolar and hyperfine couplings of the U-$5f$ electrons to the muon at the two stopping sites. 

\begin{figure}
\centering
\includegraphics[width=8.5cm]{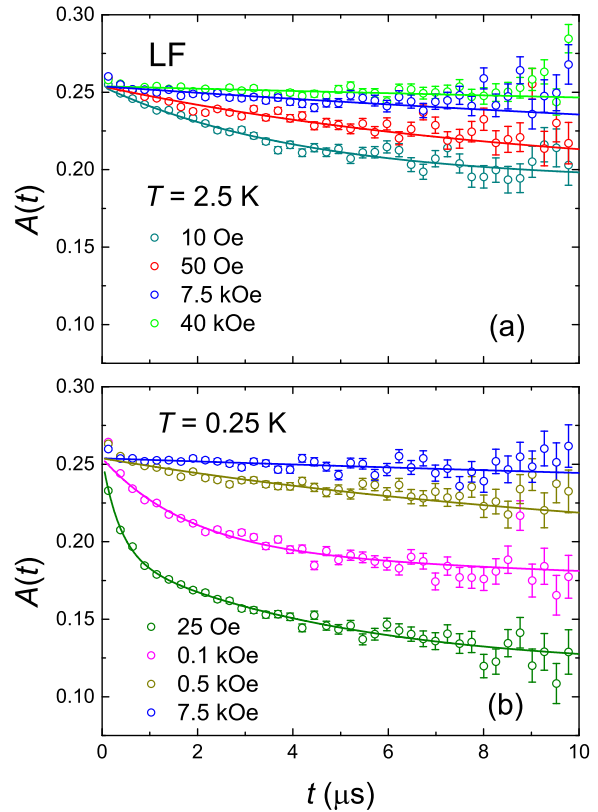}
\caption{(Color online) Representative LF-$\mu$SR asymmetry spectra at (a) 2.5~K and (b) 0.25~K, for several different values of the applied magnetic field.
The solid curves are fits to Eq.~(\ref{eq:1}).}
\label{fig2}
\end{figure}

\begin{figure}
\centering
\includegraphics[width=8.5cm]{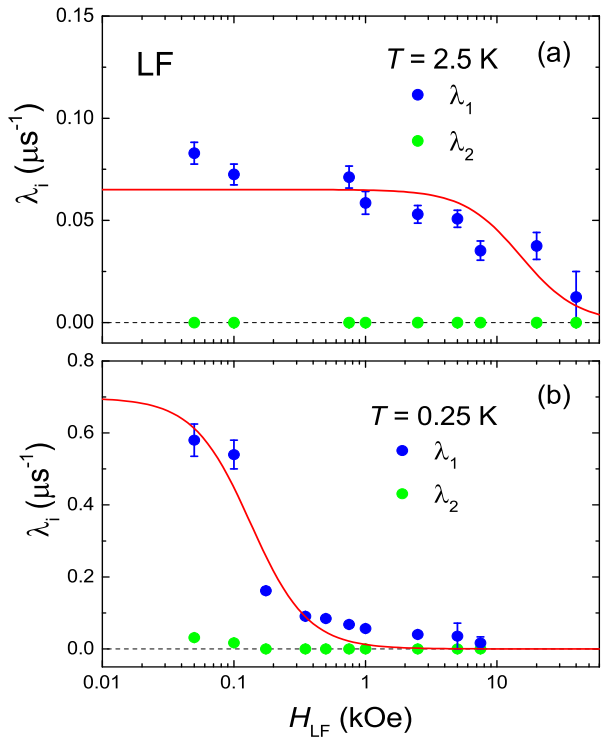}
\caption{(Color online) Field dependence of the relaxation rates $\lambda_1$ and $\lambda_2$ from the fits of the LF-$\mu$SR asymmetry spectra at (a) 2.5~K, and (b) 0.25~K.
The solid red curves are fits of $\lambda_1(H_{\rm LF})$ to Eq.~(\ref{eqn:Redfield}).} 
\label{fig3}
\end{figure}

To confirm the dynamic nature of the magnetism, LF-$\mu$SR measurements were performed for various longitudinal applied fields $H_{\rm LF}$. 
Representative LF-$\mu$SR asymmetry spectra for $T \! = \! 2.5$~K and 0.25~K are shown in Fig.~\ref{fig2}. The LF signals are reasonably described by Eq.~(\ref{eq:1}). Figure~\ref{fig3} shows the dependence of the fitted relaxation rates $\lambda_1$ and $\lambda_2$ on $H_{\rm LF}$.
Also shown in Fig.~\ref{fig3} are fits of the field dependence of the larger relaxation rate $\lambda_1$ to the Redfield equation \cite{Schenck:1985}       
\begin{equation}
\lambda_1(H_{\rm LF}) \! = \! \frac{\lambda_1(H_{\rm LF} = 0)}{1 + \left(\gamma_\mu H_{\rm LF} \tau \right)^2} \, ,
\label{eqn:Redfield}
\end{equation}
where $\lambda_1(H_{\rm LF} \! = \! 0) \! = \! 2 \gamma_\mu^2 \langle B_{\rm loc}^2 \rangle \tau$, $\langle B_{\rm loc}^2 \rangle$ is the mean of the 
square of the transverse components of the time-varying local magnetic field at the muon site, and $\tau$ is the characteristic fluctuation
time. The fit for 2.5~K yields $\lambda_1(H_{\rm LF} = 0) \! = \! 0.065(5)$~$\mu$s$^{-1}$, $\tau \! = \! 8(3) \times 10^{-10}$~s and $B_{\rm loc} \! = \! 76(22)$~G, 
whereas the fit for 0.25K yields $\lambda_1(H_{\rm LF} = 0) \! = \! 0.70(9)$~$\mu$s$^{-1}$, $\tau \! = \! 9(2) \times 10^{-8}$~s and $B_{\rm loc} \! = \! 23(4)$~G.   
We could not confirm similar fluctuation rates at the second muon site, because $\lambda_2$ is much smaller and not well resolved for most fields. 

\begin{figure}
\centering
\includegraphics[width=8.5cm]{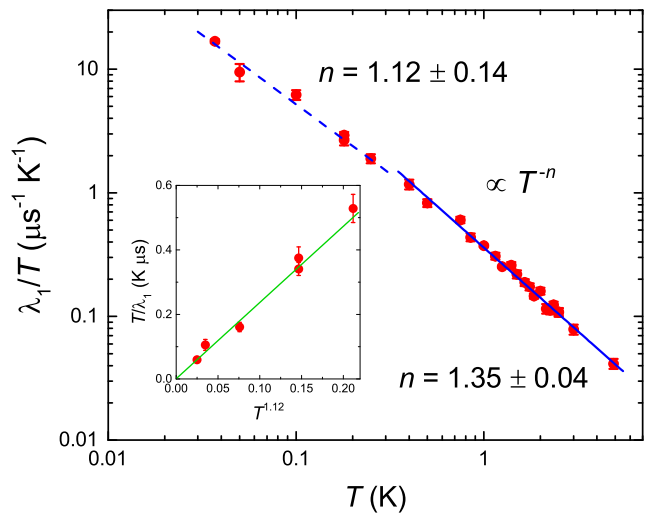}
\caption{(Color online) Temperature dependence of $\lambda_1/T$ ($\equiv \! 1/T_1T$) for zero field. The solid blue line is a fit of the
data over $0.4 \! \leq \! T \! \leq \! 4.9$~K to the power-law equation $1/T_1T \! \propto \! T^{-n}$, which yields the exponent $n \! = \! 1.35 \! \pm \! 0.04$. 
The dashed line is a similar fit over $0.037 \! \leq \! T \! \leq \! 0.3$~K, yielding $n \! = \! 1.12 \! \pm \! 0.14$. The inset
shows a plot of $T/\lambda_1$ (($\equiv \! T_1T$) versus $T^{1.12}$ with a linear fit that yields the $T \! = \! 0$ intercept 
$T/\lambda_1 \! = \! (0.7 \! \pm \! 4.2) \! \times \! 10^{-3}$~K~$\mu$s.} 
\label{fig4}
\end{figure}

Above $\sim \! 150$~K, the magnetic susceptibility $\chi(T)$ of UTe$_2$ is described by a Curie-Weiss law with an effective magnetic moment $\mu_{\rm eff}$ that is
close to the expected value ($3.6 \! \mu_{\rm B}$/U) for localized U-$5f$ electrons and a Weiss temperature $\theta \! \sim \! -100$~K \cite{Ikeda:06}. 
However near $\! \sim \! 35$~K, $\chi(T)$ for $H \! \parallel \! b$-axis exhibits a maximum that suggests the U-$5f$ electrons may become more itinerant at lower temperatures.   
Figure~\ref{fig4} shows the temperature dependence of $\lambda_1/T$, where $\lambda_1$ ($\equiv \! 1/T_1$) is the larger of the two dynamic ZF exponential 
relaxation rates. The phenomenological self-consistent renormalization (SCR) theory for intinerant ferromagnetism \cite{Moriya:91}, predicts 
that $1/T_1T \! \propto \! T^{-4/3}$ near a FM quantum critical point (QCP) in a three-dimensional metal \cite{Ishigaki:96}. As shown in Fig.~\ref{fig4}, this
behavior is observed down to $T \! = \! 0.4$~K. The deviation below $\sim \! 0.3$~K suggests a breakdown in SCR theory close to the presumed FM QCP.
The inset of Fig.~\ref{fig4} shows that $T_1T$ (which is proportional to the inverse of the imaginary part of the dynamical local spin susceptibility) 
goes to zero as $T \! \rightarrow \! 0$, which provides evidence for the ground state of UTe$_2$ being close to a FM QCP. 
     
\begin{figure}
\centering
\includegraphics[width=8cm]{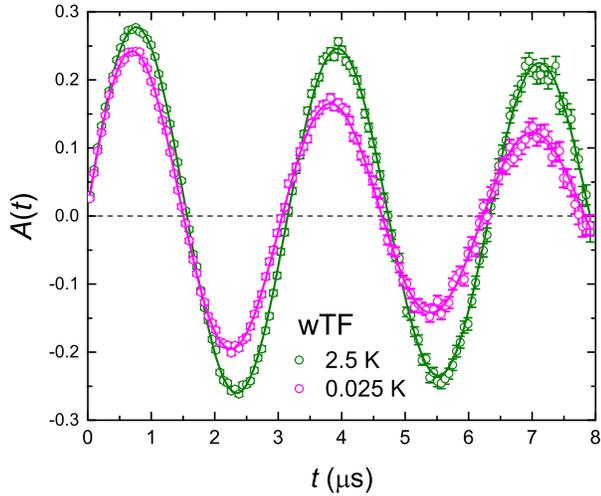}
\caption{(Color online) Weak TF-$\mu$SR asymmetry spectra recorded for $H \! = \! 23$~Oe. The solid curves are fits to Eq.~(\ref{eq:2}).}
\label{fig5}
\end{figure}

Figure~\ref{fig5} shows wTF-$\mu$SR asymmetry spectra above and far below $T_c$. 
The data were fit to the following sum of two exponentially-damped precessing terms due to the sample and an undamped temperature-independent 
precessing component due to muons that missed the sample  
\begin{eqnarray} 
A(t) = A(0)P_z(t) = & \cos & (2 \pi \nu t + \phi) \displaystyle\sum_{i = 1}^{2} A_i e^{-\Lambda_i t} \nonumber \\
& + & A_{\rm B} e^{-\Delta^2 t^2} \cos(2 \pi \nu_{\rm B} t + \phi) \, ,
\label{eq:2}
\end{eqnarray} 
where $\phi$ is the initial phase of the muon spin polarization {\bf P}(0) relative to the $x$-direction. 
The fits yield $A_1 \! + \! A_2 \! = \! 0.176(4)$ and $0.165(4)$ for $T \! = \! 2.5$~K and 0.025~K, respectively. The lower-critical field $H_{c1}(T)$ of UTe$_2$
is unknown, but presumably quite small. The smaller value of $A_{\rm S}$ at 0.025~K may be due to partial flux expulsion, if
$H_{c1}(T \! = \! 0.025~$K) is somewhat larger than the applied 23~Oe local field. Regardless, the small difference between $A_{\rm S}$ at the two temperatures 
indicates that the magnetic volume sensed by the muon above and below $T_c$ is essentially the same. Consequently, the superconductivity must reside in
spatial regions where there are magnetic fluctuations.

\begin{figure}
\centering
\includegraphics[width=8.7cm]{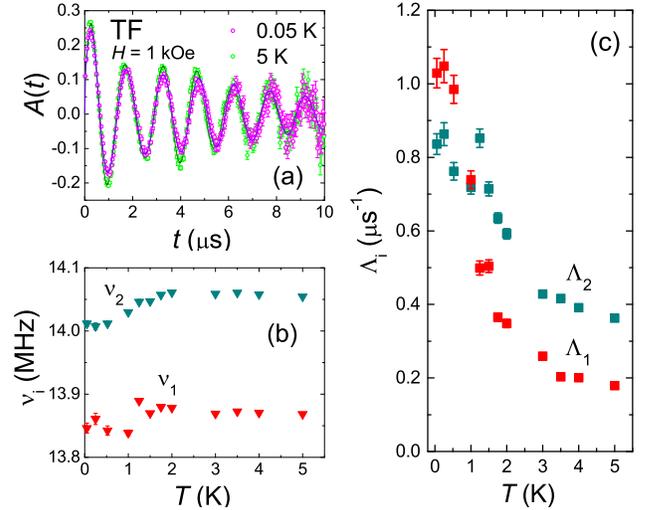}
\caption{(Color online) (a) TF-$\mu$SR asymmetry spectra at $T \! = \! 0.05$~K and 5~K for a magnetic field $H \! = \! 1$~kOe applied parallel 
to the $z$-direction, displayed in a rotating reference frame frequency of 13.15~MHz. The solid curves are fits to Eq.~(\ref{eq:4}).
Temperature dependence of the fitted (b) muon spin precession frequencies, and (c) TF relaxation rates.}
\label{fig6}
\end{figure}

Figure~\ref{fig6}(a) shows reprensentative TF-$\mu$SR asymmetry spectra recorded for $H \! = \! 1$~kOe. Once again the TF signals were fit to the sum
\begin{eqnarray} 
A(t) = A(0)P_x(t) = &  \displaystyle\sum_{i = 1}^{2} & A_i e^{-\Lambda_i t} \cos(2 \pi \nu_i t + \psi) \nonumber \\
& + & A_{\rm B} e^{-\Delta^2 t^2} \cos(2 \pi \nu_{\rm B} t + \psi) \, , 
\label{eq:4}
\end{eqnarray}
where $\psi$ is the initial phase of the muon spin polarization {\bf P}(0) relative to the $z$-direction.
The exponentially-damped terms account for muons stopping at the two sites in the sample, and the Gaussian-damped term accounts for muons that missed
the sample. The precession frequencies $\nu_i$ are a measure of the local field $B_{\mu, i}$ sensed by the muon at the two stopping sites, 
where $\nu_i \! = \! (\gamma_\mu / 2 \pi) B_{\mu, i}$ and $\gamma_\mu/ 2\pi$ is the muon gyromagnetic ratio.
The applied 1~kOe field induces a polarization of the U-$5f$ moments and a corresponding relative muon frequency shift (Knight shift),
which is different for the two muon sites. Fits of the TF asymmetry spectra to Eq.~(\ref{eq:4}) were performed assuming the background term 
is independent of temperature, and the ratio of the asymmetries $A_1$, $A_2$ and $A_{\rm B}$ are the same as determined from the analysis of 
the ZF asymmetry spectra. 

The temperature dependence of $\nu_1$ and $\nu_2$ are shown in Fig.~\ref{fig6}(b). Below $T \! \sim \! 1.6$~K there is a decrease 
in $\nu_1$ and $\nu_2$ compatible with the estimate of $\sim \! 0.2$~\% for the SC diamagnetic shift from the
relation \cite{Abrikosov:57} $-4 \pi M \! = \! (H_{c2} \! - \! H)/[1.18(2 \kappa^2 \! - \! 1) \! + \! n]$, with $H_{c2} \! = \! 200$~kOe, 
$H \! = \! 1$~kOe, $\kappa \! = \! 200$, and $n \! \leq \! 1$. 
However, the temperature dependence of the TF relaxation rates $\Lambda_1$ and $\Lambda_2$ [see Fig.~\ref{fig6}(c)] do not exhibit a significant 
change in behavior at $T_c$. This indicates that $\Lambda_1$ and $\Lambda_2$ are dominated by the internal magnetic field distribution associated with 
the magnetic fluctuations and the London penetration depth $\lambda_{\rm L}$ is quite long --- as is the case for other 
uranium-based superconductors in which $\lambda_{\rm L} \! \gtrsim \! 10,000$~\AA \cite{Gross:89}. The magnetic fluctutions may also contribute to $\nu_1(T)$ 
and $\nu_2(T)$ by adding or opposing the SC diamagnetic shift. Interestingly, the SC diamagnetic shift is not
observed in the NMR Knight shift data for a powder sample \cite{Ran:18}, although this may be a consequence of anisotropic averaging of the
NMR interactions. 
 
In conclusion, we observe a gradual slowing down of magnetic fluctuations with decreasing temperature below 5~K, consistent with weak
FM fluctuations approaching a magnetic instability. However, we find no evidence for magnetic order down 
to 0.025~K. Hence there is no phase transition to FM order in UTe$_2$ preceding or coinciding with the onset of superconductivity.
The magnetic volume fraction is not significantly reduced below $T_c$, indicating that the superconductivity coexists with the fluctuating magnetism.
Lastly, we note that because the relaxation rate of the ZF-$\mu$SR signal below 5~K is dominated by dynamic local fields, it is not possible to
determine whether spontaneous static magnetic fields occur below $T_c$ due to time-reversal symmetry breaking in the SC state.
  
\begin{acknowledgments}
We thank the staff of TRIUMF's Centre for Molecular and Materials Science for technical assistance. J.E.S. acknowledges support from Natural Sciences 
and Engineering Research Council of Canada. Research at the University of Maryland was supported by the Department of Energy, Office of
Basic Energy Sciences under Award No. DE-SC0019154, and the Gordon and Betty Moore Foundation's
EPiQS Initiative through Grant No. GBMF4419. S.R.S acknowledges support from the National Institute of Standards and Technology Cooperative Agreement 70NANB17H301.
\end{acknowledgments}


\end{document}